\documentclass[aps,prb,superscriptaddress,reprint,longbibliography,floatfix]{revtex4-1}
\usepackage[english]{babel}
\usepackage{amsmath}
\usepackage{amssymb}
\usepackage[separate-uncertainty = true]{siunitx}
\usepackage{graphicx}
\usepackage{xcolor}
\usepackage{booktabs}

\begin{document}
\title{Heisenberg Exchange and Dzyaloshinskii-Moriya Interaction in Ultrathin CoFeB Single and Multilayers} 
\date{\today}
\author{Tobias B\"ottcher}
\affiliation{Fachbereich Physik and Landesforschungszentrum OPTIMAS, Technische Universit\"at Kaiserslautern, Gottlieb-Daimler-Stra\ss e 46, 67663 Kaiserslautern, Germany}
\affiliation{Graduate School of Excellence Materials Science in Mainz, Staudingerweg 9, 55128 Mainz, Germany}

\author{Kyujoon Lee}
\affiliation{Institut f\"ur Physik, Johannes Gutenberg-Universit\"at, Staudingerweg 7, 55128 Mainz, Germany}

\author{Frank Heussner}
\affiliation{Fachbereich Physik and Landesforschungszentrum OPTIMAS, Technische Universit\"at Kaiserslautern, Gottlieb-Daimler-Stra\ss e 46, 67663 Kaiserslautern, Germany}

\author{Samridh Jaiswal}
\affiliation{Institut f\"ur Physik, Johannes Gutenberg-Universit\"at, Staudingerweg 7, 55128 Mainz, Germany}
\affiliation{Singulus Technologies AG, Hanauer Landstra\ss e 103, 63796 Kahl am Main, Germany}

\author{Gerhard Jakob}
\affiliation{Institut f\"ur Physik, Johannes Gutenberg-Universit\"at, Staudingerweg 7, 55128 Mainz, Germany}

\author{Mathias Kl\"aui}
\affiliation{Institut f\"ur Physik, Johannes Gutenberg-Universit\"at, Staudingerweg 7, 55128 Mainz, Germany}
\affiliation{Graduate School of Excellence Materials Science in Mainz, Staudingerweg 9, 55128 Mainz, Germany}

\author{Burkard Hillebrands}
\affiliation{Fachbereich Physik and Landesforschungszentrum OPTIMAS, Technische Universit\"at Kaiserslautern, Gottlieb-Daimler-Stra\ss e 46, 67663 Kaiserslautern, Germany}

\author{Thomas Br\"acher}
\affiliation{Fachbereich Physik and Landesforschungszentrum OPTIMAS, Technische Universit\"at Kaiserslautern, Gottlieb-Daimler-Stra\ss e 46, 67663 Kaiserslautern, Germany}

\author{Philipp Pirro}
\affiliation{Fachbereich Physik and Landesforschungszentrum OPTIMAS, Technische Universit\"at Kaiserslautern, Gottlieb-Daimler-Stra\ss e 46, 67663 Kaiserslautern, Germany}
\begin{abstract} 

We present results of the analysis of Brillouin Light Scattering (BLS) measurements of spin waves performed on ultrathin single and multirepeat CoFeB layers with adjacent heavy metal layers. From a detailed study of the spin-wave dispersion relation, we independently extract the Heisenberg exchange interaction (also referred to as symmetric exchange interaction), the Dzyaloshinskii-Moriya interaction (DMI, also referred to as antisymmetric exchange interaction), and the anisotropy field. We find a large DMI in CoFeB thin films adjacent to a Pt layer and nearly vanishing DMI for CoFeB films adjacent to a W layer. Furthermore, the residual influence of the dipolar interaction on the dispersion relation and on the evaluation of the Heisenberg exchange parameter is demonstrated. In addition, an experimental analysis of the DMI on the spin-wave lifetime is presented. All these parameters play a crucial role in designing skyrmionic or spin-orbitronic devices.

\end{abstract} 
\maketitle

\section*{Introduction}
The remarkable pace of development of CMOS-based information processing devices seen over the past decades can very well be described by Moore's law \cite{Moore.1965,Keyes.2006}. However, this progress is based on a continuous decrease of device size \cite{Waldrop.2016} which, in connection with Joule heating, results in an approach towards critical power densities. In this context, magnetic objects such as skyrmions and spin waves are envisaged to form the basis of a new generation of information storage and processing devices \cite{Parkin.2008,Fert.2013,Jiang.2015,Kang.2016,Zazvorka.2019,Zhang.2020}.

For the stabilization of skyrmions, not only the symmetric Heisenberg exchange interaction but also the presence of an antisymmetric exchange contribution favoring a chiral alignment of spins is a crucial requirement \cite{MoreauLuchaire.2016}. As predicted by Dzyaloshinskii and Moriya \cite{Dzyaloshinskii.1958,Moriya.1960}, low-symmetry systems can exhibit such a contribution to the exchange interaction. Mediated by a broken inversion symmetry and the selection of a capping material with a large spin-orbit interaction \cite{Fert.1980}, the Dzyaloshinskii-Moriya interaction (DMI) can be especially pronounced in ultrathin magnetic films adjacent to a layer of a heavy metal. In this case, it is referred to as interfacial DMI (iDMI) \cite{Stashkevich.2015}.

The DMI strength in single layers can be measured using asymmetric domain expansion \cite{Balk.2017}, asymmetric switching of triangles \cite{Han.2016}, or stripe domain annihilation \cite{Woo.2016,Soucaille.2016,Jaiswal.2017}. However, such methods require an additional determination or estimation of the symmetric exchange interaction since the strength of the iDMI cannot be determined independent from this parameter. In contrast, an investigation of spin waves allows for an independent determination of symmetric as well as the antisymmetric exchange interaction. In general, spin waves constitute a powerful tool for the characterization of magnetic thin films, micro-, and nanostructures, and they possess a spatial chirality making them sensitive to the presence of DMI \cite{Moon.2013}. DMI and other material parameters can be traced by the characterization of the thermally populated spin-wave dispersion relation using Brillouin Light Scattering (BLS) spectroscopy \cite{Nembach.2015,Belmeguenai.2015,Di.2015,Chaurasiya.2016}.

At the same time, spin waves \cite{Bloch.1930} and their corresponding quasi-particles, magnons \cite{Serga.2010,Kruglyak.2010,Chumak.2015}, have been employed in many prototype devices such as in a magnon transistor \cite{Chumak.2014}, in spin-wave majority gates \cite{Klingler.2014,Klingler.2015,Ganzhorn.2016b,Kanazawa.2017,Fischer.2017}, and in many others \cite{Chumak.2009,Schultheiss.2014,Nikitin.2015,Heussner.2017,Wang.2018,Meyer.2018b}. In this context, non-reciprocal spin-wave propagation as a consequence of DMI might also constitute an interesting tool to boost the capabilities of spin-wave logic devices \cite{Bracher.2014}.

Among the materials composing the thin films under investigation, for spin-orbitronics devices heavy metal/CoFeB bilayers are very relevant. CoFeB (in various compositions) plays an important role in many spintronic applications such as MRAM \cite{Stamps.2014} and devices based on the propagation of spin waves \cite{Yu.2012,Heussner.2018}. Its properties can be widely tuned by annealing \cite{Conca.2014}, and it is easy to handle using sputtering techniques, rendering this alloy important for many applications. Heavy metals like Pt and W show a large spin-orbit coupling which, for example, leads to a large spin Hall angle \cite{Liu.2011,Hoffmann.2013} which is desirable for devices involving spin orbit torques \cite{Garello.2013b}.

In this work, we present measurements of the DMI strength in single-repeat and multirepeat systems of both Pt/CoFeB/MgO and W/CoFeB/MgO stacks using BLS spectroscopy. In particular, we employ wave vector-resolved BLS spectroscopy resulting in a direct measurement of the spin-wave dispersion relation. Considering the role of the symmetric and antisymmetric exchange interaction on the dispersion relation \cite{Kalinikos.1986,Moon.2013}, in contrast to an analysis of domain expansion, BLS spectroscopy allows for an independent determination of both exchange contributions.
The symmetric exchange of both systems is found to be identical within the measurement errors whereas we find a pronounced iDMI in the Pt-based stack and nearly vanishing iDMI in the W-based system.

\section*{Samples}
The investigated single-repeat samples consist of an underlayer (UL) of either Pt or W sputter-deposited onto a thermally oxidized Si substrate using a \textit{Singulus Rotaris} deposition system. The full single-repeat stack is UL(\SI{5}{nm})/\linebreak[1]{}Co$_{20}$Fe$_{60}$B$_{20}$(\SI{0.6}{nm})/\linebreak[1]{}MgO(\SI{2}{nm})/\linebreak[1]{}Ta(\SI{5}{nm}) and it has been investigated as deposited. 
The corresponding multirepeat samples investigated in this work consist of ten repetitions of UL(\SI{5}{nm})/\linebreak[1]{}Co$_{20}$Fe$_{60}$B$_{20}$(\SI{0.6}{nm})/\linebreak[1]{}MgO(\SI{2}{nm}), deposited onto a thermally oxidized Si substrate and finally capped with a $\SI{5}{nm}$ Ta layer using the same sputtering system.

The value of the saturation magnetization used in all calculations has been obtained by vibrating sample magnetometry (VSM) from a W(\SI{5}{nm})/\linebreak[2]{}Co$_{20}$Fe$_{60}$B$_{20}$(\SI{0.6}{nm})/\linebreak[2]{}MgO(\SI{2}{nm})/\linebreak[2]{}Ta(\SI{5}{nm}) stack on an oxidized Si substrate. From this, we obtain $M_\mathrm{S}=\SI{1388}{kA \per m}$.

\section*{Extraction of exchange constants from the spin-wave dispersion relation}
For in-plane magnetized thin films, spin waves with wave vector $\vec{k}$ propagating perpendicularly to the static magnetization $\vec{M}$ exhibit a spatial chirality depending on the relative orientation of $\vec{k}$ and $\vec{M}$. Consequently, the presence of iDMI results in an increase or decrease of the spin-wave frequency depending on the propagation direction, making spin waves a useful probe in the investigation of iDMI.

For spin waves propagating in in-plane saturated ultrathin films, the corresponding influence of the iDMI on the dispersion relation \cite{Kalinikos.1986} can be described by a frequency shift linear in the spin-wave wave vector \cite{Moon.2013}. Together with the remaining terms of the dispersion relation for spin waves in films with uniaxial out-of-plane anisotropy, the spin-wave frequency $f_\mathrm{SW}$ is given by the relation \cite{Di.2015}

\begin{widetext}
\begin{equation}
f_\mathrm{SW}(k) = \frac{\gamma \mu_0}{2 \pi}\sqrt{ \Big(  H_\mathrm{ext} + \lambda_\mathrm{ex}k^2 + M_\mathrm{S} \mathrm{g}(kt) \Big)
  \cdot \Big( H_\mathrm{ext} - H_\mathrm{U} + \lambda_\mathrm{ex}k^2 + M_\mathrm{S} - M_\mathrm{S} \mathrm{g}(kt) \Big)  } 
+ \frac{\gamma}{\pi M_\mathrm{S}}D \cdot k
\label{Eq:Dispersion_full}
\end{equation}
\end{widetext}

with the gyromagnetic ratio $\gamma$ (here, $\gamma = \SI{176}{rad~ T^{-1} ns^{-1}}$ is assumed), the permeability of vacuum $\mu_0$, the film thickness $t$, and the uniaxial anisotropy field $H_\mathrm{U}$ connected to the anisotropy constant $K_\mathrm{U} $ via $\mu_0 H_\mathrm{U} = 2K_\mathrm{U} / M_\mathrm{S}$.
The parameter $D$ describes the DMI strength \cite{Moon.2013}, and the influence of the symmetric exchange interaction is included in the exchange stiffness $\lambda_\mathrm{ex}=2A / (\mu_0 M_\mathrm{S})$ with the exchange constant $A$.
Furthermore, the dipole-dipole interaction is represented by the function
\begin{equation}
\mathrm{g}(x) = 1- \left[ 1- \exp (-|x|) \right] / |x| 
\label{Eq:g}
\end{equation}
with $x=kt$ a dimensionless parameter.
%
%

In case the DMI is purely of  interfacial origin, the DMI constant $D$ can be related to the film thickness resulting in an interfacial DMI constant \cite{Belmeguenai.2015}
\begin{equation}
D_\mathrm{S}=Dt.
\end{equation}

For the measurement of the spin-wave dispersion relation, BLS spectroscopy is employed, which is based on the inelastic scattering of photons with magnons. For this process, momentum as well as energy conservation laws hold \cite{Sebastian.2015}:
\begin{align}
\hbar \vec{k}_\mathrm{out} &= \hbar \vec{k}_\mathrm{in} \pm \hbar \vec{k}_\mathrm{SW} \\
\hbar \omega_\mathrm{out} &= \hbar \omega_\mathrm{in} \pm \hbar \omega_\mathrm{SW}.
\end{align}

Here, (in) and (out) describe the wave vector and frequency of the incident and scattered photon, respectively, and the label (SW) relates to the spin-wave wave vector and frequency, respectively. A positive sign in the above equations holds for the case of an anti-Stokes process in which a magnon is annihilated, whereas the negative sign holds for the case of a Stokes process in which a magnon is created.

By operating the setup in the backscattering geometry \cite{Sebastian.2015}, wave-vector resolved probing of spin waves and, with that, a direct measurement of the spin-wave dispersion relation is possible. 
Considering the influence of the DMI on the dispersion relation (cf.~eq.~\ref{Eq:Dispersion_full}), such experiments have proven to be a valuable method to get an insight into properties of samples under the influence of DMI \cite{Kim.2019b}.

The DMI constant can be extracted from the linear slope of the shift of the spin-wave frequency under reversal of the spin-wave wave vector:
\begin{equation}
f(k)_\mathrm{asym}=\frac{|f(-k)_\mathrm{Stokes}-f(k)_\mathrm{Stokes}|}{2} = \frac{\gamma}{\pi M_\mathrm{S}}D k.
\label{Eq:Asymmetric_f}
\end{equation}
The same equation holds for the anti-Stokes signal.
In this context, we would like to point out that a reversal of the sign of the external field, by symmetry arguments, corresponds to a reversal of the experimentally probed spin-wave wave vector and an inversion of Stokes and anti-Stokes signal. With that, any experimental offset error in the frequency measurement can be minimized.

At the same time, in order to more reliably extract other sample parameters from the dispersion relation, it can be beneficial to symmetrize the spin-wave frequency according to the relation

\begin{widetext}
\begin{equation}
f(k)_\mathrm{sym} = \frac{|f(k)_\mathrm{Stokes}|+|f(k)_\mathrm{anti-Stokes}|}{2}
= \frac{\gamma \mu_0}{2 \pi} \sqrt{ \Big(  H_\mathrm{ext} + \lambda_\mathrm{ex}k^2 + M_\mathrm{S} \mathrm{g}(kt) \Big)
  \cdot \Big( H_\mathrm{ext} - H_\mathrm{U} + \lambda_\mathrm{ex}k^2 + M_\mathrm{S} - M_\mathrm{S} \mathrm{g}(kt) \Big)  }
\label{Eq:Symmetric_f}  
\end{equation}
\end{widetext}

which is independent of the frequency shift induced by the antisymmetric exchange interaction.

Please note in this context that the given dispersion relation neglects any inhomogeneity of the magnetic parameters over the film thickness such as an inhomogeneous saturation magnetization or different surface anisotropies on the upper and lower surface which, principally, could also lead to a frequency non-reciprocity \cite{Hillebrands.1990}. However, as shown in \textcite{Gladii.2016}, these effects are very small in the present case because of the ultrathin film thickness.

In the experiment, we use a laser with a wavelength of $\lambda = \SI{532}{nm}$. Hence, the spin-wave wave vector probed is $k = 4 \pi \sin(\varphi) / \lambda $ with the angle of incidence $\varphi$. The probing laser beam is incident perpendicular to the applied field which lies in the sample plane. Thus, in case the film is saturated, magnetostatic surface spin waves are probed which propagate perpendicularly to the magnetization ($\vec{k} \perp \vec{M}$).
Since no microwave field for the excitation of spin waves is applied in the investigation, solely thermal spin waves are detected in the experiments.

\section*{Results}

\subsection*{Pt/CoFeB/MgO}

\begin{figure}
	 \centering
	 \includegraphics[height=6.3cm]{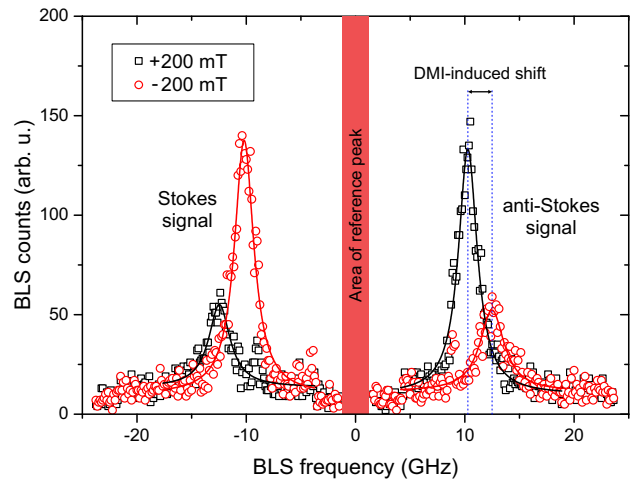}
	\caption{~(color online)~BLS spectra obtained from the single repeat Pt/CoFeB/MgO sample at an angle of light incidence of $\varphi=\SI{60}{\degree}$ for positive and negative applied field with $|\mu_0 H_\mathrm{ext}|=\SI{200}{mT}$. The respective spin-wave frequencies are obtained from the center frequencies of Lorentzian peak functions fitted to the signal peaks. The DMI-induced frequency shift under field reversal, i.e., a reversal of the sign of the spin-wave wave vector, is marked in the spectrum. Also, a field direction-dependent asymmetry between Stokes and anti-Stokes intensity is visible as expected for thin films of absorptive materials \cite{Camley.1982}.}
	\label{Fig1}
\end{figure}

In order to investigate the influence of the DMI on the spin-wave dispersion relation experimentally, an applied field of $\mu_0 H_\mathrm{ext}=\SI{\pm 200}{mT}$ is chosen. As an example, Fig.~\ref{Fig1} shows the BLS spectra recorded from the single repeat sample at an angle of incidence of $\varphi=\SI{60}{\degree}$ for positive and negative magnetic fields ($\mu_0 H_\mathrm{ext}=\pm \SI{200}{mT}$). The asymmetry of the intensity of Stokes and anti-Stokes signal might be a consequence of helicity-dependent contributions to the scattering cross section \cite{Camley.1982} but it is not of relevance for the extraction of the magnetic parameters. Further measurements at various angles of incidence are performed, thus, probing a large range of spin-wave wave vectors. The center frequency of the Stokes and anti-Stokes signal are extracted by fitting a Lorentzian peak function to the respective signals in the recorded BLS spectra.

\begin{figure*}
	 \centering
	 \includegraphics[width=\textwidth]{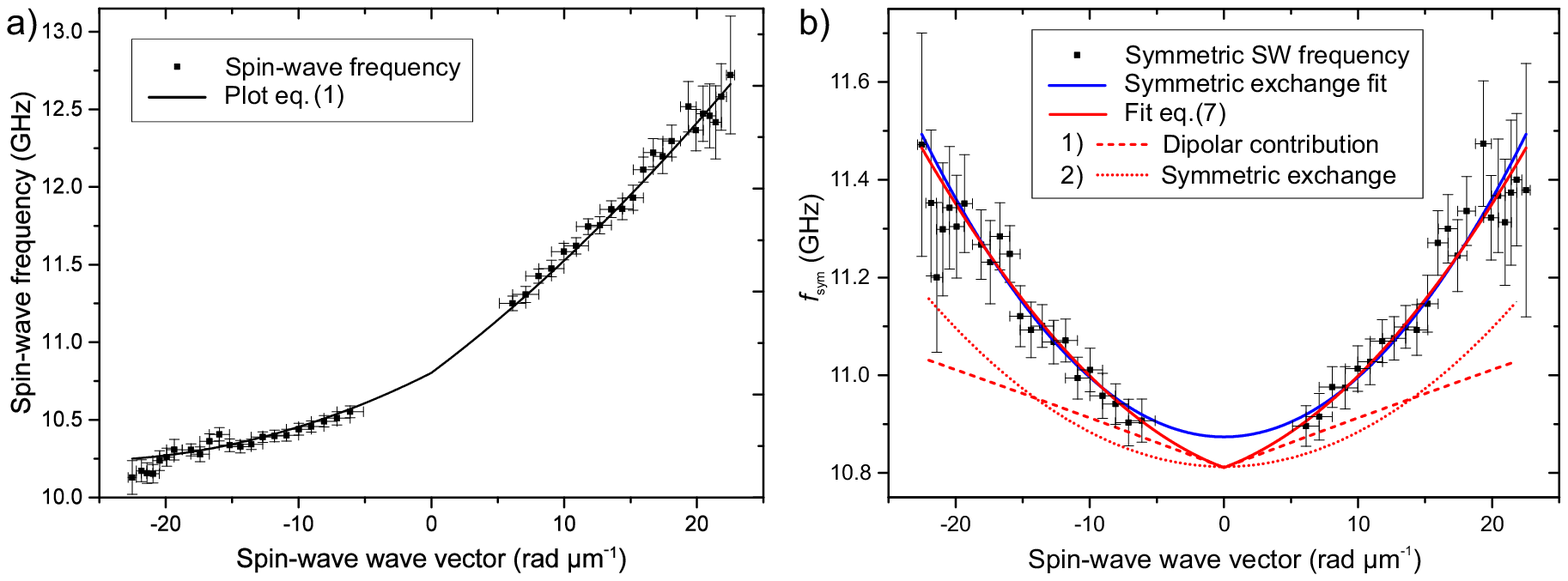}
	\caption{~(color online)~a)~Measured spin-wave frequency (black dots) and analytical model of the full dispersion relation (cf.~eq.~\ref{Eq:Dispersion_full}). For the model, the symmetric exchange constant obtained from the fit in Fig.~\ref{Fig2}b) and the DMI constant obtained from the fits in Fig.~\ref{Fig3} have been used. b)~Symmetrized spin-wave dispersion relation (black dots) and a fit only including symmetric exchange interaction (blue curve). The red solid line is a fit of eq.~\ref{Eq:Symmetric_f} resulting in $A = \SI{17.60 \pm 3.14}{pJ \per m}$. The red dashed line is the dipolar contribution and red dotted line is the (symmetric) exchange contribution to the dispersion, respectively.}
	\label{Fig2}
\end{figure*}

The full dispersion relation is shown by the black squares in Fig.~\ref{Fig2}a). Clearly, it features a pronounced asymmetry with respect to wave-vector inversion as a consequence of the DMI. This asymmetry can be used to obtain the DMI constant according to eq.~\ref{Eq:Asymmetric_f}. Furthermore, we would like to point out that the group velocity, which is the derivative of the dispersion relation with respect to the wave vector, is positive in the entire probed wave vector range. Consequently, the group velocity, or, in other words, the magnon energy flow, is unidirectional for all spin-wave wave vectors probed in the experiment. This effect, which is due to a comparably strong iDMI might provide interesting opportunities for the application of such layer systems in spin-wave logic devices \cite{GarciaSanchez.2014,Fert.2013,Bracher.2017}.

\begin{figure}
	 \centering
	 \includegraphics[height=6.3cm]{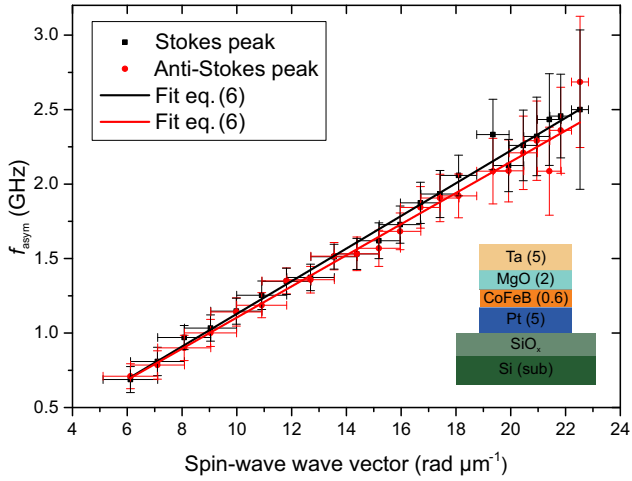}
	\caption{~(color online)~Difference in the peak position of the Stokes and the anti-Stokes peak under reversal of the spin-wave wave vector. The measured values clearly exhibit a linear dependence which is in accordance with the DMI-induced frequency shift to the dispersion relation being linearly dependent on the wave vector. We find $D = +\SI{1.33 \pm 0.09}{mJ \per m^2}$. CoFeB stands for the composition Co$_{20}$Fe$_{60}$B$_{20}$.}
	\label{Fig3}
\end{figure}

The black line in Fig.~\ref{Fig2}a) shows the full dispersion relation including the DMI contribution as given in eq.~\ref{Eq:Dispersion_full}. 
For the plot, the external field ($\mu_0 H_\mathrm{ext} = \SI{200}{mT}$), the magnetic film thickness ($t = \SI{0.6}{nm}$), as well as the saturation magnetization ($M_\mathrm{S} = \SI{1388}{kA \per m}$) have been used. The symmetric exchange constant has been taken from the fit in Fig.~\ref{Fig2}b) and the DMI constant is the one extracted from the data shown in Fig.~\ref{Fig3}.

The symmetrized spin-wave frequency is shown in Fig.~\ref{Fig2}b) by the black squares. Fitting the symmetrized dispersion relation as in eq.~\ref{Eq:Symmetric_f} to the data results in a symmetric exchange constant of $A = \SI{17.60 \pm 3.14}{pJ \per m}$. At this point, we would like to underline the fact that, even though the magnetic film thickness is only $\SI{0.6}{nm}$, neglecting the dipolar interaction significantly falsifies this analysis. This shall be illustrated by the additional plots in Fig.~\ref{Fig2}b). 

First, the blue curve is a fit of a model which only includes a symmetric exchange contribution quadratic in the spin-wave wave vector and the FMR frequency as an offset, i.e., $f_\mathrm{SW} = f_\mathrm{FMR} + \lambda_\mathrm{ex} k^2$. This would lead to $A = \SI{30.26 \pm 3.93}{pJ \per m}$ which drastically overestimates the actual value of the symmetric exchange constant. 

Second, for an adequate analysis of the results, the importance of the dipolar interaction is illustrated better by a separate presentation of the dipolar contribution. It is shown by the dashed red line in Fig.~\ref{Fig2}b). This curve is a plot of the dispersion relation with the parameters obtained from the fit of the full dispersion relation but with the symmetric exchange constant $A$ set to zero. The dipolar contribution is linear in the spin-wave wave vector to very good approximation and is definitely of importance despite the very small film thickness. For the maximum spin-wave wave vector probed ($k_\mathrm{max}$), it amounts to approximately $\SI{200}{MHz}$. Thus, with $f_\mathrm{SW}(k_\mathrm{max})-f_\mathrm{FMR} \approx \SI{600}{MHz}$, the dipolar contribution amounts to roughly one third of this shift.

Again, the dotted red curve in Fig.~\ref{Fig2}b) is the above-mentioned parabolic spin-wave dispersion with the exchange constant as extracted from the fit of the full dispersion relation. It underlines, on the one hand, the fact that the parabolic model for the symmetrized dispersion relation is an incomplete approach and, on the other hand, that the contribution of the dipolar interaction to the dispersion relation cannot be neglected. Please note in this context, that the full dispersion relation is not the sum of the dipolar and exchange contribution but it is more complex as visible from eq.~\ref{Eq:Dispersion_full}.

The influence of the antisymmetric exchange interaction, i.e., the DMI, is extracted according to eq.~\ref{Eq:Asymmetric_f}. These data, which are presented in Fig.~\ref{Fig3}, show the frequency difference of the spin waves under reversal of the sign of the wave vector which linearly increases with the absolute value of the wave vector.
With the above-mentioned value for $M_\mathrm{S}$ and with the film thickness of $t=\SI{0.6}{nm}$, the values for the DMI constant obtained from the Stokes data and the anti-Stokes data agree very well within their respective error bars. We obtain

\begin{align}
\nonumber
D &= +\SI{1.33 \pm 0.09}{mJ \per m^2}   \nonumber \\
D_\mathrm{S} &= +\SI{0.80 \pm 0.06}{pJ \per m}.  \nonumber
\end{align}

Ma \textit{et al.} \cite{Ma.2018} found a DMI constant of $|D_\mathrm{S}|=\SI{0.97}{pJ \per m}$ using BLS for a CoFeB/Pt interface with the same CoFeB composition as for the samples under investigation in this work.
This result is in reasonable agreement with our findings.

Additionally, we find $\mu_0 H_\mathrm{U} = \SI{1.198 \pm 0.038}{T}$ which is smaller than $\mu_0 M_\mathrm{S} = \SI{1.744}{T}$ confirming the in-plane anisotropy of the film under investigation.



\subsection*{W/CoFeB/MgO}

Similar measurements as for the Pt/CoFeB/MgO film have been performed for the W/CoFeB/MgO film. From a field-sweep BLS measurement (not shown) in which a softening can be observed \cite{Stamps.1991}, it is found that this film features a perpendicular magnetic anisotropy (PMA). Measurements are performed at an applied field strength of $|\mu_0 H_\mathrm{ext}| = \SI{400}{mT}$ such that the film is saturated in its plane.

\begin{figure}
	 \centering
	 \includegraphics[height=6.3cm]{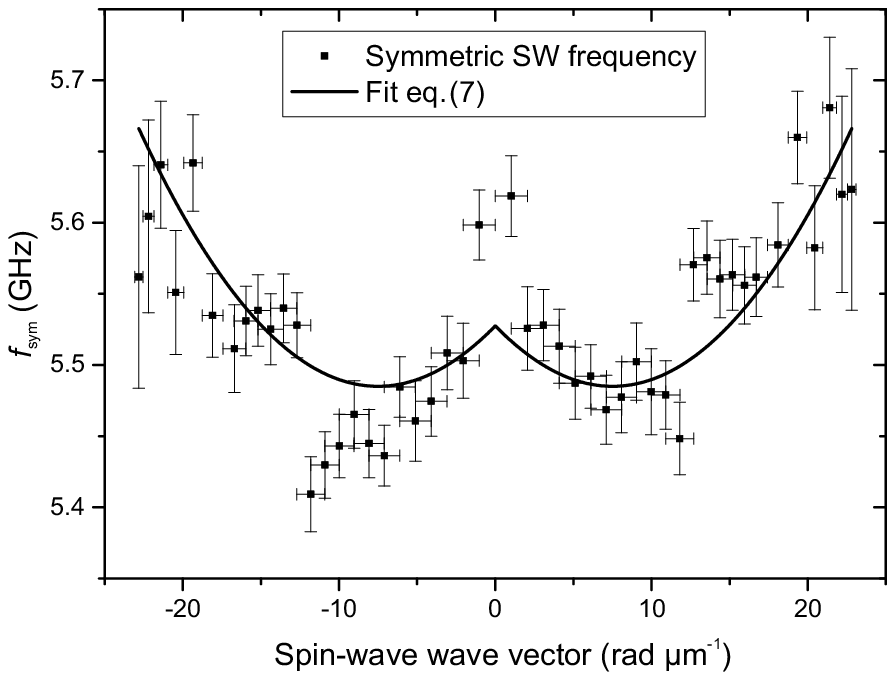}
	\caption{~(color online)~Symmetrized spin-wave dispersion relation measured at an externally applied magnetic field  strength of $\SI{400}{mT}$ in the W/CoFeB/MgO sample. The black solid line is a fit of the dispersion relation according to eq.~\ref{Eq:Dispersion_full}. From the fit we obtain $A = \SI{15.00 \pm 2.82}{pJ \per m}$. The flat character in the wave-vector range covered is consequence of the interplay between PMA, symmetric exchange and dipolar interaction.}
	\label{Fig4}
\end{figure}

The symmetrized dispersion relation extracted from BLS measurements performed with the W/CoFeB/MgO sample is shown in Fig.~\ref{Fig4}. Its remarkable flatness is a consequence of the interplay between symmetric exchange, the dipolar interaction, and the PMA causing a backward curvature which is clearly visible in the corresponding fit curve depicted by the solid black line in Fig.~\ref{Fig4}. 
For the fit, we again kept the saturation magnetization constant. The film thickness is  $t = \SI{0.6}{nm}$ and the applied magnetic field is $\mu_0 H_\mathrm{ext} = \SI{400}{mT}$. With that, we obtain $A = \SI{15.00 \pm 2.82}{pJ \per m}$ for the symmetric exchange constant which is in good agreement with the value found for Pt/CoFeB/MgO. For the uniaxial anisotropy field we find $\mu_0 H_\mathrm{U} = \SI{2.047 \pm 0.001}{T}$. The value $\mu_0 H_\mathrm{sat} = \mu_0 H_\mathrm{U} - \mu_0 M_\mathrm{S} = \SI{302}{mT}$ corresponds well to the value of the applied field at which the maximum softening has been observed in the field-sweep measurement (not shown). Hence, the first remarkable difference when compared to the Pt/CoFeB/MgO sample is the presence of a significantly larger PMA in the W-based stack.

\begin{figure}
	 \centering
	 \includegraphics[height=6.3cm]{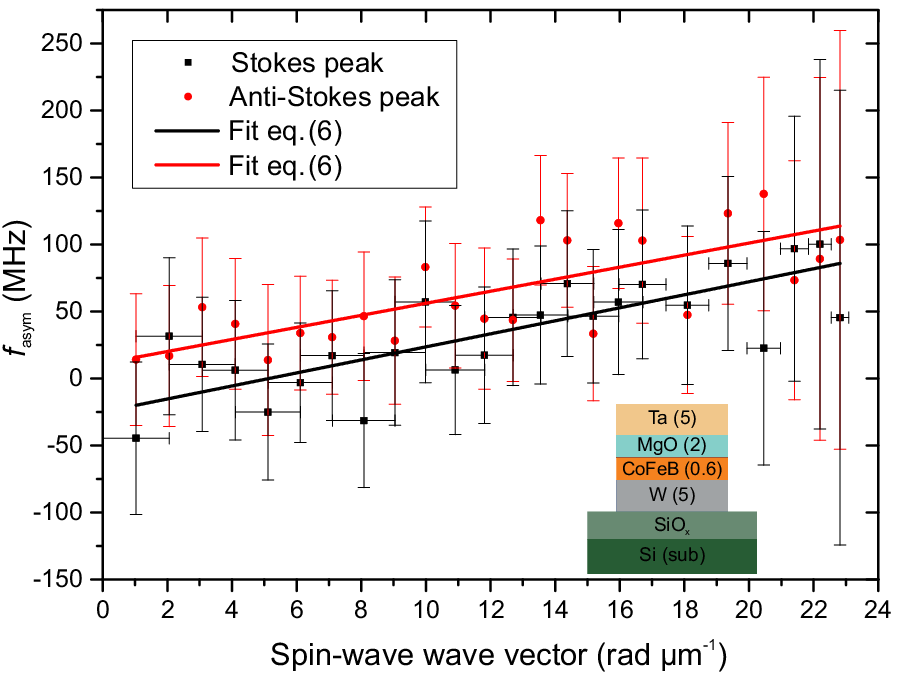}
	\caption{~(color online)~Difference in the peak position of the Stokes and the anti-Stokes peak under reversal of the spin-wave wave vector for the W/CoFeB/MgO sample. From the linear fits we find an average DMI constant of $D = +\SI{0.06 \pm 0.03}{mJ \per m^2}$. CoFeB stands for the composition Co$_{20}$Fe$_{60}$B$_{20}$.}
	\label{Fig5}
\end{figure}

The DMI constant is again obtained by analyzing the frequency shift $f_\mathrm{asym}$ under reversal of the spin-wave wave vector.
The corresponding data is shown in Fig.~\ref{Fig5} together with linear fits used to evaluate the DMI constant. It turns out that the frequency shift is significantly smaller than that observed for the Pt-based stack. Both the values for the DMI constant obtained from the Stokes and anti-Stokes data, respectively, agree very well and the obtained values are 

\begin{align}
D &= +\SI{0.06 \pm 0.03}{mJ \per m^2}   \nonumber  \\
D_\mathrm{S} &= +\SI{0.04 \pm 0.02}{pJ \per m}. \nonumber
\end{align}

Thus, the DMI is much weaker in the W-based thin film than in the Pt-based one even though both Pt and W are elements with a large spin-orbit coupling. Hence, both elements could be expected to induce a pronounced iDMI in the adjacent magnetic layers \cite{Fert.1980,Chaurasiya.2016}. However, this seems not to be the case for the W-based stack investigated in our experiments. Possible reasons might be a dependence of the strength of the induced interfacial DMI on the phase of the tungsten layer \cite{Demasius.2016}.

\subsection*{Multirepeat samples} 
Besides single layers, also magnetic multilayers are of large interest, in particular, since they improve the stability of skyrmions \cite{MoreauLuchaire.2016}. 

\begin{figure}
	 \centering
	 \includegraphics[height=6.3cm]{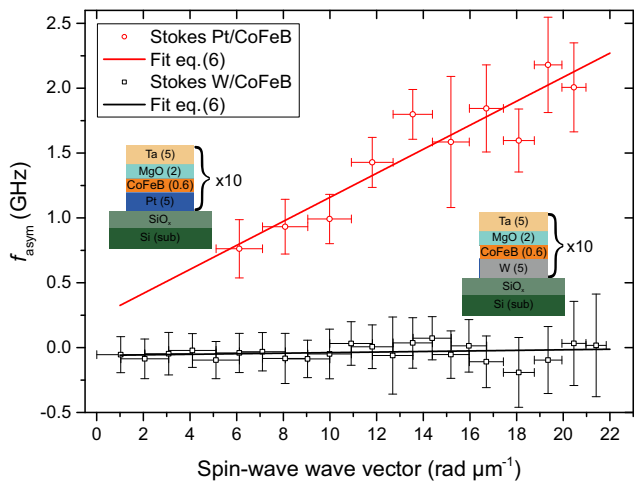}
	\caption{~(color online)~Frequency difference of the Stokes signal in the multirepeat samples under reversal of the external magnetic field. The linear fit to the data allows for an estimation of the DMI constant. We find $D =  \SI{1.15 \pm 0.57}{mJ \per m^2}$ for the Pt/CoFeB-based multilayer and $D =  \SI{0.03 \pm 0.08}{mJ \per m^2}$ for the W/CoFeB-based multilayer, respectively. CoFeB stands for the composition Co$_{20}$Fe$_{60}$B$_{20}$.}
	\label{Fig6}
\end{figure}

We note that in the multirepeat samples under investigation, we observe a significantly weaker BLS signal as compared to the single layer resulting in a larger uncertainty of the obtained sample properties. We choose an externally applied magnetic field of $\mu_0 H_\mathrm{ext}=\SI{\pm 350}{mT}$ for the Pt/CoFeB-based multilayer and $\mu_0 H_\mathrm{ext}=\SI{\pm 400}{mT}$ for the W/CoFeB-based multilayer to perform wave-vector resolved measurements according to the same procedure as described above for the investigation of the single layers. From an  analysis of the Stokes signal similar to the single repeats, we obtain the results depicted in Fig.~\ref{Fig6}. 

Again, we find a linear increase in the difference between the peak positions under a reversal of the field with an increase of the spin-wave wave vector. Fitting eq. \ref{Eq:Asymmetric_f} to the data obtained from the Pt/CoFeB-based multilayer, we find a DMI constant of 
\begin{align}
D &=  +\SI{1.15 \pm 0.57}{mJ \per m^2}  \nonumber 
\end{align}
which is in agreement with the DMI strength found for a single stack of Pt/CoFeB/MgO within the error bars. This result is in accordance with the assumption that the individual magnetic layers of the multirepeat stack are exchange decoupled and have the same properties as the magnetic layer in the single stack samples.

The same is true for the multirepeat stack based on W/CoFeB. In this case, we find a DMI constant of
\begin{align}
D &=  +\SI{0.03 \pm 0.08}{mJ \per m^2}   \nonumber
\end{align}
which again is in agreement with the findings for the single repeat sample.

\begin{table*}
\caption{\label{tab:summary}Summary of the sample properties obtained in this work. For all films, $M_\mathrm{S} = \SI{1388}{kA \per m}$ has been assumed. Numbers in brackets denote layer thickness in $\SI{}{nm}$.}
\begin{ruledtabular}
\begin{tabular}{ccccc}
  & $D~(\SI{}{mJ \per m^2})$ & $D_\mathrm{S}~(\SI{}{pJ \per m})$ & $A~(\SI{}{pJ \per m})$ &  $H_\mathrm{U}~(\SI{}{T})$ \\
\midrule 
Pt(5)/Co$_{20}$Fe$_{60}$B$_{20}$(0.6)/MgO(2)/Ta(5)  & $+\SI{1.33 \pm 0.09}{}$ & $+\SI{0.797 \pm 0.056}{}$ & $\SI{17.60 \pm 3.14}{}$ & $\SI{1.198 \pm 0.038}{}$ \\
$[$Pt(5)/Co$_{20}$Fe$_{60}$B$_{20}$(0.6)/MgO(2)$]_{10}$/Ta(5) & $+\SI{1.15 \pm 0.57}{}$ & - & - & -\\
W(5)/Co$_{20}$Fe$_{60}$B$_{20}$(0.6)/MgO(2)/Ta(5) & $+\SI{0.06 \pm 0.03}{}$ & $+\SI{0.035 \pm 0.020}{}$ & $\SI{15.00 \pm 2.82}{}$ & $\SI{2.047 \pm 0.001}{}$ \\
$[$W(5)/Co$_{20}$Fe$_{60}$B$_{20}$(0.6)/MgO(2)$]_{10}$/Ta(5) & $+\SI{0.03 \pm 0.08}{}$ & - & - & -\\
\end{tabular}
\end{ruledtabular}
\end{table*}

\subsection*{Spin-wave lifetime}
A parameter which can give further insight into the material properties such as the damping is the lifetime $\tau$ of a spin-wave mode with wave vector $k$, which in the case of thin films is significantly influenced by the heavy metal layer due to spin pumping \cite{Tserkovnyak.2005} and two-magnon scattering at interface imperfections and, in addition, has been reported to depend on DMI \cite{Zakeri.2012}. In concrete terms, DMI is expected to cause a difference of the linewidth between the Stokes and the anti-Stokes signal.

Via the phenomenological loss theory \cite{Stancil.2009}, the lifetime can be connected to the spin-wave dispersion relation \cite{Kalinikos.1986}. As given in the work of \textcite{Bracher.2017}, in the absence of DMI and for thin films with spin waves propagating perpendicularly to the magnetization direction, the lifetime can be expressed as
\begin{equation}
\tau_0 = \frac{1}{2 \pi \alpha} \left[ \frac{\gamma \mu_0}{2 \pi} \left( H_\mathrm{ext} + \lambda_\mathrm{ex}k^2 - \frac{H_\mathrm{U}}{2}+ \frac{M_\mathrm{S}}{2} \right) \right] ^{-1}
\label{Eq:Lifetime}
\end{equation}
with the effective Gilbert damping parameter $\alpha$. In the BLS measurements, provided the signal linewidth is large compared to the frequency resolution of the setup ($\approx \SI{100}{MHz}$), the spin-wave lifetime is directly linked to the signal linewidth $\delta  \! f$ via the relation
\begin{equation}
\delta \! f = \frac{1}{4 \pi \tau}.
\label{Eq:Linewidth}
\end{equation}
The presence of iDMI leads to a modification of the spin-wave lifetime. With the DMI-induced frequency difference $f_\mathrm{asym}$ (cf.~eq.~\ref{Eq:Asymmetric_f}) between Stokes and anti-Stokes signals, the lifetime of the respective counterpropagating waves (plus and minus sign, respectively) is found to be
\begin{equation}
\tau_\pm  = \frac{\tau_0}{1 \pm \frac{f_\mathrm{asym}}{f_\mathrm{sym}}}
\label{Eq:Lifetime_DMI}
\end{equation}
with the frequencies $f_\mathrm{asym}$ and $f_\mathrm{sym}$ as defined in eqs.~\ref{Eq:Asymmetric_f} and \ref{Eq:Symmetric_f}.

\begin{figure*}
	 \centering
	 \includegraphics[height=6.5cm]{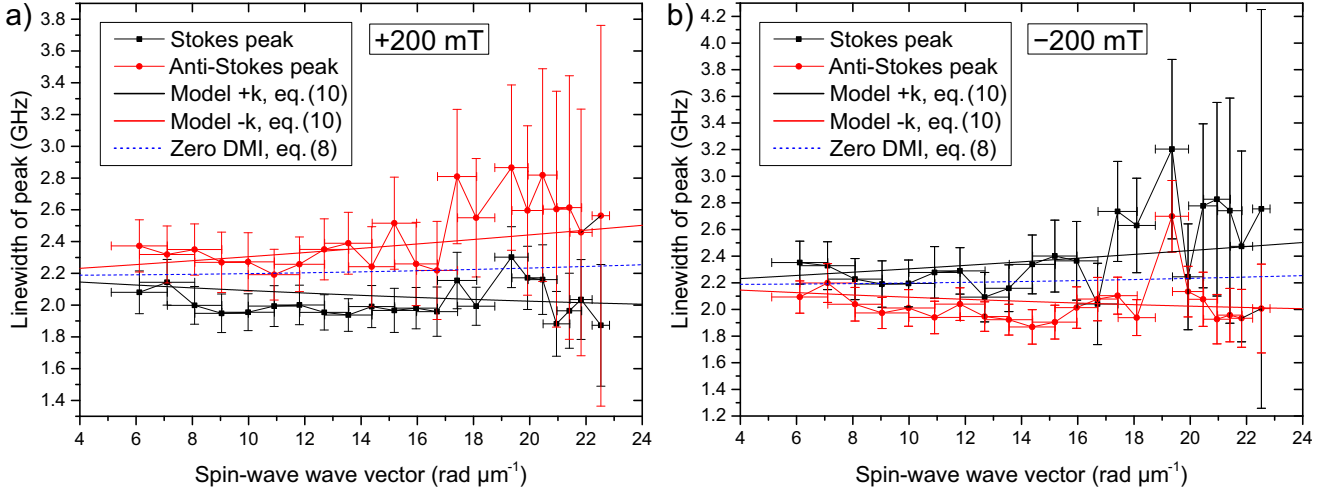}
	\caption{~(color online)~Signal linewidth of the Stokes (black squares) and the anti-Stokes (red circles) peak as a function of $k$ for a) $\mu_0 H_\mathrm{ext}=\SI{+200}{mT}$ and b) $\mu_0 H_\mathrm{ext}=\SI{-200}{mT}$. Solid lines are calculated spin-wave lifetimes for non-zero DMI and both propagation directions (black and red, respectively) and with $D=0$ (blue). The Gilbert damping is set to $\alpha=\SI{0.33}{}$. Lines between experimental values are guides-to-the-eye.}
	\label{Fig7}
\end{figure*}

The signal linewidth of Stokes and anti-Stokes signal are shown in Figs.~\ref{Fig7}a) and b). The model curves are the calculated lifetimes according to eqs.~\ref{Eq:Lifetime} and \ref{Eq:Lifetime_DMI} with an effective Gilbert damping parameter of $\alpha=\SI{0.33}{}$ and $D = \SI{+1.33}{mJ \per m^2}$. This value for the Gilbert damping parameter is reasonable given the fact that the film under investigation is ultrathin and that an adjacent Pt layer is known to significantly enhance the Gilbert damping due to spin pumping effects \cite{Tserkovnyak.2002,RuizCalaforra.2015}.
In order to quantify this effect, we estimate the contribution of spin pumping to the total Gilbert damping. It can be expressed as \cite{Conca.2016b}

\begin{equation}
\alpha_\mathrm{SP} = \frac{\gamma \hbar}{4 \pi M_\mathrm{S} t} g_\mathrm{eff}^{\uparrow\downarrow}
\label{Eq:Gilbert_SP}
\end{equation}

with $\hbar = h/(2 \pi)$ where $h$ is Planck's constant, and the spin mixing conductance $g_\mathrm{eff}^{\uparrow\downarrow}$.
For the spin mixing conductance, a value of \SI{4d-19}{m^{-2}} is being reported for a Co/Pt interface \cite{Zhang.2015c} as well as for a Co$_{20}$Fe$_{60}$B$_{20}$/Pt interface \cite{RuizCalaforra.2015}. With this, we can estimate the spin pumping damping enhancement to be $\alpha_\mathrm{SP}=\SI{0.23}{}$. Accordingly, the damping of the CoFeB layer, which is not caused by spin pumping effects is about $\alpha_\mathrm{int}=\SI{0.10}{}$. Thus, although we can only estimate the different contributions, we can state that spin pumping is likely the dominating contribution to the overall damping.

Concerning the influence of the DMI on the linewidth, we can state that, as expected, the asymmetry of the linewidths of the Stokes and the anti-Stokes peaks inverts under field, i.e., wave-vector reversal. Furthermore, an increase of the linewidth towards higher spin-wave wave vectors is visible. Thus, the systematic wave-vector-dependent influence of the DMI on the spin-wave linewidth can be clearly observed and is well in line with results found elsewhere \cite{Di.2015b,Chaurasiya.2016}.

\section*{Conclusion}

In summary, from measurements of the thermal spin-wave spectrum, we independently determined the DMI constant and the symmetric exchange constant in ultrathin CoFeB layers deposited on Pt and W, respectively. We find a strong DMI induced by the adjacent Pt layer, whereas the W-based stack exhibits a surprisingly small interfacial DMI constant. In addition, we show that the contribution of the dipolar interaction to the dispersion relation cannot be neglected, especially in the context of an accurate extraction of the symmetric exchange constant. 

The parameters found for the multirepeat stacks are in good agreement with the ones found for the single repeat samples underlining the assumption of exchange-decoupled magnetic layers in the multilayer samples. The influence of DMI on the spin-wave properties can be also traced by an analysis of the spin-wave lifetime.

\subsection*{Acknowledgements}
This work has been funded by the Deutsche Forschungsgemeinschaft (DFG, German Research Foundation) by TRR 173 - 268565370 ("Spin+X", Projects A01 and B01) and by SPP 2137 ("Skyrmionics", Projects 403512431 and 403502522).

\end{document}